# A Fast Compressive Sensing Based Digital Image Encryption Technique using Structurally Random Matrices and Arnold Transform

Nitin Rawat[1*], Pavel Ni[2], Rajesh Kumar[3]

Nitin Rawat
Dept. of Information & Communication
Gwangju Institute of Science & Technology, Gwangju, South Korea
E-mail: nitin@gist.ac.kr
Phone: +821032921985

Pavel Ni
Gwangju Institute of Science & Technology, Gwangju, South Korea

Rajesh Kumar
Jaypee University of Information Technology, Waknaghat, Solan, India

**Abstract**

A new digital image encryption method based on fast compressed sensing approach using structurally random matrices and Arnold transform is proposed. Considering the natural images to be compressed in any domain, the fast compressed sensing based approach saves computational time, increases the quality of the image and reduces the dimension of the digital image by choosing even 25 % of the measurements. First, dimension reduction is utilized to compress the digital image with scrambling effect. Second, Arnold transformation is used to give the reduced digital image into more complex form. Then, the complex image is again encrypted by double random phase encoding process embedded with a host image; two random keys with fractional Fourier transform are been used as a secret keys. At the receiver, the decryption process is recovered by using TwIST algorithm. Experimental results including peak-to-peak signal-to-noise ratio between the original and reconstructed image are shown to analyze the validity of this technique and demonstrated our proposed method to be secure, fast, complex and robust.

## 1. Introduction

With the rapid growth of information security, cryptography has become an essential research subject where several encryption techniques have been realized to expand the area of cryptography [1–5]. Among them, the double random phase encryption (DRPE) has widely known for its simple implementation, robustness, and easy application on image formats i.e. black & white, gray level [6]. The DRPE secures the contents of the data by scrambling, which can be unlocked only by using the right key. It can be extracted in several domains, such as Fourier transform (FT) [7], Fresnel transform (FrT) [3] and fractional Fourier transform (FRFT) [8,9]. An FRFT gives greater complexity to the system by giving an extra parameter of the transform order from 0 to 1[8,9], which enlarges the key space resulting in a better and secure data protection as compared to the Fourier transformation (FT) [7].

Compressive sensing (CS) technique has drawn attention in the field of signal processing & image acquisition [10,11] and has shown a dramatic change in the encryption methods as well [12,13]. Recently, an encryption method based on encryption is proposed using gray [12] and color images [13]. Although, it takes a lot of computational time and the recovery is poor when reducing the number of measurements.

In this paper, we offer a novel image encryption technique based on CS, which can save the computational time and recover the data by using only 25 % of the measurements from the original image as compared with the conventional CS based encryption techniques used in encryption [12,13]. Meanwhile, in order to further enhance the security and complexity of the system, a well-known Arnold transform (AT) and DRPE techniques are used. Numerical results show our system to be more complex, secure and can resist various attacks.

## 2. Fundamental Knowledge

### 2.1. Double random phase encoding using fractional Fourier transform

An optical setup using a 4f system in Fig. 1 describes the encoding process using the DRPE. In the system, the image $I(x,y)$ to be encrypted is multiplied by the independent phase functions and is transformed through FRFT order. $\theta_0(x,y)$ and $\psi_0(u,v)$ represents the two key functions in the space and frequency domain respectively. The FRFT techniques are complex compared to the conventional FT since FRFT provides an extra degree of freedom, where $(\alpha,\beta) = p\pi/2$ are the angles at which FRFT can be calculated.

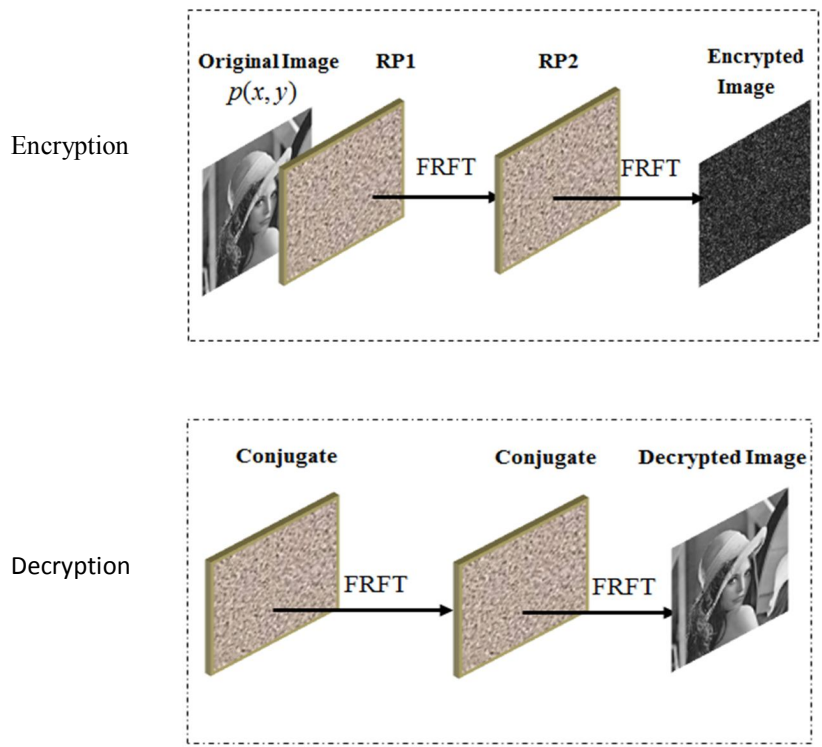

**Fig. 1.** Conventional double random phase encoding using fractional Fourier transform.

The FRFT of a function $f_i$ of $a^{th}$ order is expressed as:

$$f_i\{I(x,y)\} = \int\int_{-\infty}^{+\infty} K_i(x,y;x',y')I(x,y)dxdy \quad \ldots\ldots (1)$$

where $K_i$ is the transform kernel expressed as:

$$K_i(x,y;x',y') = A_\varphi \exp[i\pi(x^2+y^2+x'^2+y'^2)\cot\varphi_\alpha - 2(xx'+yy')\cot\varphi_\alpha] \quad \ldots\ldots (2)$$

where $A_\varphi = \exp[-i\pi\,\text{sgn}(\sin\varphi)/4 + i\varphi/2]$, $\varphi = \alpha\pi/2$ is the transform angle and $0 < |\alpha| < 2$ is the range. The encoding process can be expressed as:

$$E(x,y) = F_\alpha\{F_\beta\{[I(x,y)\exp[j2\pi\theta(x,y)]\exp[j2\pi\omega(u,v)]\}\} \quad \ldots\ldots (3)$$

where $F_{\alpha,\beta}^{-1}$ represents the inverse FRFT through the order of $(\alpha,\beta)$ and $x$ and $y$ are the coordinates of space domain, $u$ and $v$ are the coordinates of FRFT domain, respectively. The decoding process can be expressed as:

$$I(x,y) = F_\beta^{-1}\{\{F_\alpha^{-1}[E(x,y)\exp[-j2\pi\theta(x,y)]\exp[-j2\pi\theta(u,v)]\}\} \quad \ldots\ldots (4)$$

where $F_{\alpha,\beta}^{-1}$ represents the inverse FRFT through the order of $(\alpha,\beta)$. $\theta_0(x,y)$ and $\psi_0(u,v)$ represents the two random phase masks which act as a key itself in the encoding and decoding process.

*2.2. Arnold transformation*

AT is a well know technique to achieve encryption algorithm in image transformation domain [14]. After the image is taken into the system, an AT is employed by changing the pixel coordinates of digital image and further scramble it several times to provide security system key. Let an original image of size $N \times N$, having coordinates $(x,y)$ is scrambled. The two-dimensional AT of the shifted coordinates $(x',y')$ is expressed as:

$$\begin{pmatrix} x' \\ y' \end{pmatrix} = \begin{pmatrix} 1 & 1 \\ 1 & 2 \end{pmatrix} \begin{pmatrix} x \\ y \end{pmatrix} (\mathrm{mod}\ N)\quad x,y \in \{1,2,.....,N-1\} \qquad \ldots\ldots (5)$$

Then its feedback equation :

$$p_{ij}^{n+1} = Ap_{ij}^{n}(\mathrm{mod}\ N),\ p_{ij}^{n} = (i,j),\ n = 0,1,...N-1 \qquad \ldots\ldots (6)$$

where $A = \begin{pmatrix} 1 & 1 \\ 1 & 2 \end{pmatrix}$ is the transformation matrix, mod denotes the modulus after division.

Through AT, the image is replaced by Eq. (5) and further the iteration process in Eq. (6) is carried out. AT provides high speed, robustness with good efficiency.

### 2.3. Compressive sensing approach

The theory of CS heavily relies on signal or image sparsity and can efficiently extract the most efficient information from small number of measurements, i.e., to minimize the collection of redundant data [4,15]. CS says that a small non-adaptive linear measurements of a compressive signal or image have enough information to reconstruct it perfectly [4,16].

The CS theory can be summarized in the following ways:

$$y = \Phi x = \Phi \Psi \alpha$$

$$y_{M \times 1} = \Phi_{M \times N} x_{N \times 1} = \Phi \Psi \alpha_{N \times 1} \qquad \ldots\ldots (7)$$

where $y$ denotes as a $M \times 1$ vector, $\Phi$ is an $M \times N$ measurement matrix and $x$ is $N \times 1$. $\alpha$ is an image and $\Psi$ is the image representation in some basis. A schematic of CS is shown in Fig. 2 where the image is sparse in some domain and from fewer measurement; a signal can be easily extracted. The aim is to find a basis where the coefficient vector $\alpha$ is sparse, where only $K \ll N$ coefficients are non-zero. The sparsity $K$, is assumed to be much less than $N$. The image $x$ can only be recovered in CS if it satisfies the *Restricted Isometry Property* (RIP) [13]. One of the conditions is the sparsity and incoherence between the sensing matrix $\Phi$ and the sparsifying operator $\Psi$ expressed as.

$$\mu(\Phi,\Psi) = \sqrt{n} \cdot \max_{1 \leq k,j \leq N} \left| \langle \varphi_k, \psi_j \rangle \right| \qquad \ldots\ldots (8)$$

where $\varphi_k, \psi_j$ denotes the column vector of $\Phi$ and $\Psi$ respectively and $N$ is the length of the column vector. If $\Phi$ and $\Psi$ are highly correlated, the coherence is large, else it is small. It follows $\mu(\Phi,\Psi) \in [1, \sqrt{n}]$.

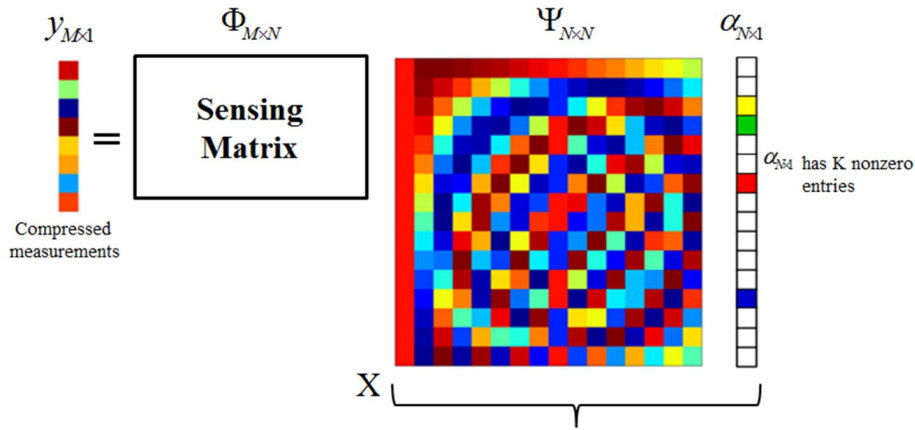

**Fig. 2.** Schematic of CS approach.

If both the sensing matrix $\Phi$ and the sparsifying operater $\Psi$ are highly uncorrelated, then a perfect signal can be easily reconstructed from small number of measurements.

The conventional least square methods can be shown to fail with high probability and thus an alternative $l_1$-norm optimization is used:

$$\hat{\alpha} = \arg\ \min \|\alpha\|_1 \quad \text{such that } y = \Phi\Psi\alpha \qquad \ldots\ldots (9)$$

$$\hat{x} = \Psi\hat{\alpha}$$

Several algorithms have been used to solve this convex optimization problem [15,16]. We have chosen the two-step iterative shrinkage threshold (TwIST) algorithm [17] which is composed of a least square minimization term.

*2.4. Fast compressive sensing approach using structurally random matrices (FSRM)*

In CS, one of the parts include the key components including the sensing matrix $\Phi$ which should be highly incoherent with the sparsifying operator $\Psi$, and the reconstruction algorithm such as basis pursuit, and iterative thresholding, orthogonal matching pursuit (OMP). Despite of achieving high incoherence and reconstructing a perfect signal, the system requires huge data memory and high computational complexity due to its completely unstructured nature [18].

Another part includes the partial Fourier which is incoherent with having sparsity in the time domain [19]. It only works well in the case where the sparsifying basis is the identity matrix [20]. These methods are sparse in specific domain, severely narrowing its scope of applications.

In this paper, we propose a fast CS based approach using structurally random matrices (FSRM) that can be sparse in any domain [20]. FSRM is able to scramble the structure of the information i.e. the image and convert the sensing matrix into a white noise like to achieve a perfect incoherent sensing. Conventional CS algorithm can recover the transform coefficient vector $\alpha$ by solving $l_1$ minimization problem [21]. For applications that require sensing operation by computing the fast transform $F$, FSRM can exploit the fast computation of matrix $F$. The computational complexity is generally spent on computing the matrix-vector multiplications $Au$ and $A^T u$ whereas in case of FSRM, the sparsifying operator $\Psi$ is fast i.e. $Au$ and $A^T u$ are fast computable.

To analysis the validity, we have use an image of size $512 \times 512$ and a well-known sparsifying basis $\Psi$ (aubechies 9/7 wavelet transform). Conventional CS method takes a lot of time whereas the scrambled FFT in wavelet domain has an optimal performance as the Fourier matrix is completely incoherent with the identity matrix which reduces the computational time.

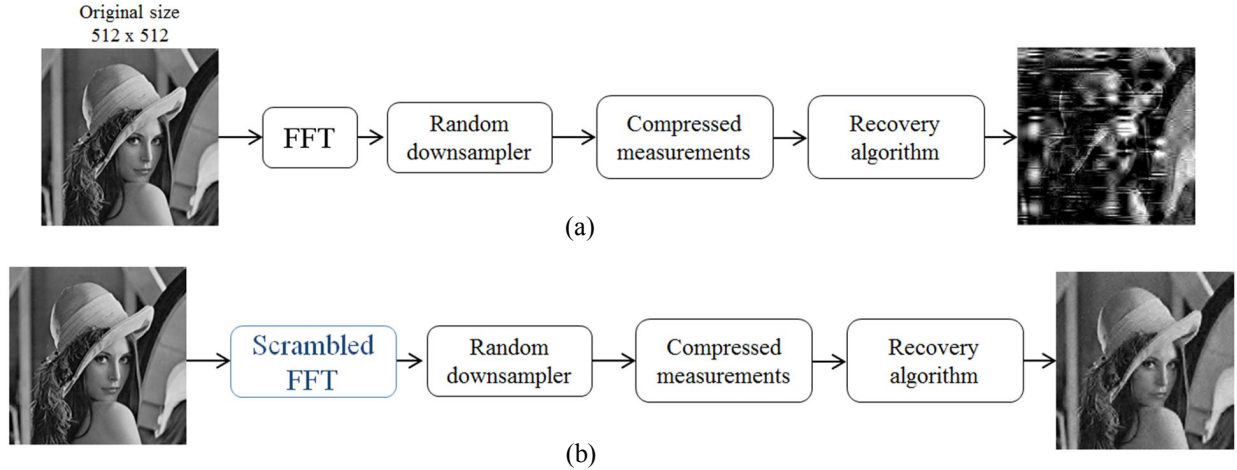

**Fig. 3.** Schematic of CS system. (a) Conventional CS system using FFT, PSNR= 16.5 dB (b) CS system using scrambled FFT, PSNR=29.4 dB.

The results are shown in Fig. 3, where the conventional and our proposed system are compared. Fig. 3(a) shows the conventional method where the FFT is used using 25 % ($M =128$) measurements of the image with a PSNR of 16.5 dB, whereas in Fig. 3(b), our system recovered the image with having a PSNR of 29.4 dB.

### 3. Digital image Encryption process

*3.1. Encryption and decryption procedure*

The flow chart of the proposed encryption method is discussed in Fig. 4 where the digital image of size 512×512 is used. CS is applied using FSRM technique where only 25 % of the measurements are used from the original image and act as first security key. The AT is applied to scramble the data and make it more complex and act as a second security key to the system. Further, the DRPE is used to get the encryption image where the two random phases act as the third and fourth key elements of the system. At the receiving channel, the original information is extracted by reversing the DRPE and HC process. Finally the information is reconstructed approximately by using the two-step iterative shrinkage threshold (TwIST) algorithm [17].

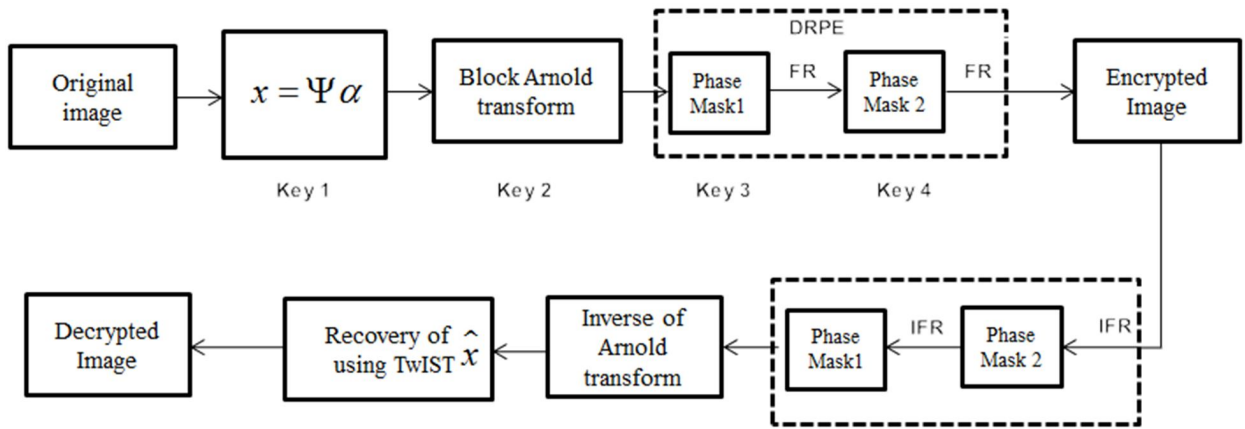

**Fig. 4.** Flow chart of the proposed system.

Without giving the sparse matrix, it is not possible to recover the exact signal. To verify the results, the peak-to-peak signal-to-noise ratio (PSNR) between the original image and decryption image are and expressed:

$$\text{PSNR}(I,I') = 10\log \frac{255^2}{\left(\frac{1}{MN}\right)\sum_{i=1}^{N}\sum_{j=1}^{M}[I(i,j)-I'(i,j)]^2} \quad \ldots (10)$$

where $I(i,j)$ and $I'(i,j)$ are the reconstructed and original image at pixel $(i,j)$ respectively.

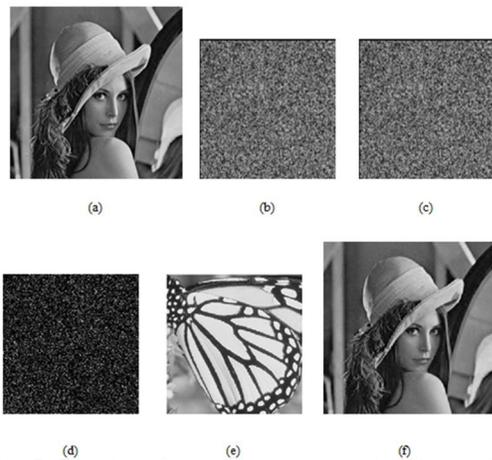

**Fig. 5.** Results of encoding process. (a) Original image, (b) scrambling image after FSRM, (c) scrambling image (d) hiding image, (e) host image, (f) final decryption image using right keys; PSNR= 28.96.

Fig. 5 shows the results of using the proposed encryption process where the PSNR calculated is 28.96 dB which indicates the good quality of an image.

## 4. The security analysis and robustness

*4.1. Robustness of the encryption image to noise and pixels cropping*

In Fig. 6, the effect of random noise is investigated where a random noise is added to a value ranged from 0 to 1. The PSNR calculated is 24.32 dB which shows the image is well recognized after applying the random noise.

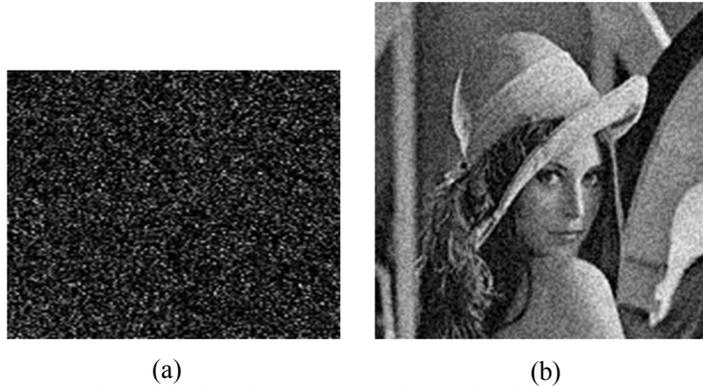

(a)            (b)

**Fig. 6.** The analysis of robustness when noise is added. (a) Encrypted image with random noise of value ranged from 0 to 1 (b) reconstruction result, PSNR= 25.2025 dB.

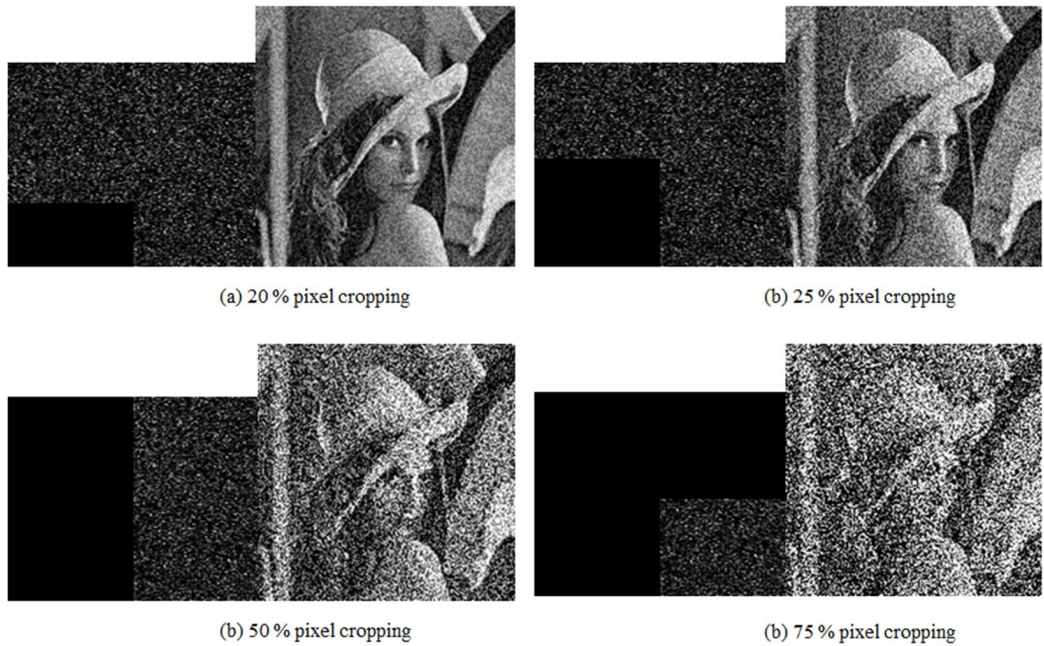

**Fig. 7.** The analysis of robustness of the encrypted image when pixel cropped, (a) 20 % pixels cropped, PSNR= 17.256 dB; (b) 25 % pixels cropped, PSNR = 15.421 dB; (c) 50 % pixels cropped, PSNR= 11.256 dB; (d) 75 % pixels cropped, PSNR= 7.052 dB.

If in case, the encrypted image is attacked by cropping the pixels; the quality of reconstruction will get worse as shown in Fig. 7. The more the pixels are cropped, the smaller the PSNR is noted which gives a valid proof of securing the image.

*4.2.Robustness to wrong key*

The FSRM based encryption method has high security and robustness. The decrypted image can be possible if and only if all the keys are available. In Fig. 9, we analyze that the decryption process is very sensitive to the right keys. Fig. 9

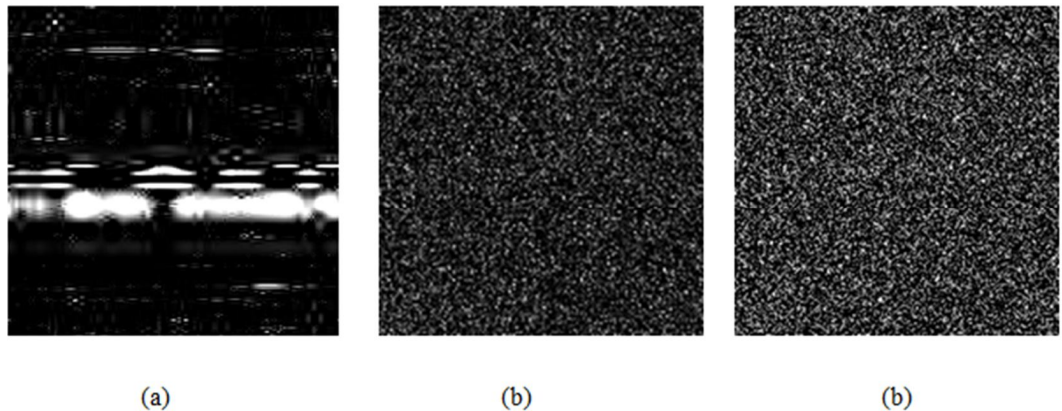

**Fig. 6.** Wrong key parameters; (a) using wrong FSRM key 1, PSNR= 3.482 dB (b) using wrong AT key 2, PSNR= 2.104 dB, (c) using wrong DRPE key 3, 4, PSNR= 1.23 dB.

The figure shows the wrong CS and FRFT parameters which are the indication to a high security encoding process.

## 5. Conclusion

In this paper, a secure digital image encryption based on fast CS technique and AT along with DPRE technique is proposed. With the help of FRSM, the size of the digital image is reduced to

25 % of the measurements with increased quality. Then the AT is applied with DRPE technique and provides a complex way represent the digital image.

Our proposed method saves the computational time, increases the quality of the image by using even 25 % of the original information and guarantees the security of the information more effectively.